# Mediating exchange bias by Verwey transition in CoO/Fe$_3$O$_4$ thin film

X. H. Liu,[1,2,a)] W. Liu,[2] Z. D. Zhang,[2] and C. F. Chang[1]
[1]*Max Planck Institute for Chemical Physics of Solids, Nöthnitzerstr. 40, 01187 Dresden, Germany*
[2]*Shenyang National Laboratory for Materials Science, Institute of Metal Research, Chinese Academy of Sciences, Shenyang 110016, China*

We report the tunability of the exchange bias effect by the first-order metal-insulator transition (known as the Verwey transition) of Fe$_3$O$_4$ in CoO (5 nm)/Fe$_3$O$_4$ (40 nm)/MgO (001) thin film. In the vicinity of the Verwey transition, the exchange bias field is substantially enhanced because of a sharp increase in magnetocrystalline anisotropy constant from high-temperature cubic to low-temperature monoclinic structure. Moreover, with respect to the Fe$_3$O$_4$ (40 nm)/MgO (001) thin film, the coercivity field of the CoO (5 nm)/Fe$_3$O$_4$ (40 nm)/MgO (001) bilayer is greatly increased for all the temperature range, which would be due to the coupling between Co spins and Fe spins across the interface. *Published by AIP Publishing.* https://doi.org/10.1063/1.5023725

## I. INTRODUCTION

Exchange bias (EB) refers to a shift in the hysteresis loop along the magnetic field axis due to the interface exchange coupling between ferromagnetic (FM) and antiferromagnetic (AFM) materials, which was first discovered by Meiklejohn and Bean in oxide-coated Co particles.[1] This phenomenon has been extensively studied because of the technological application in spintronic devices and magnetic recording.[2–4] Recently, many investigations of EB in thin films were carried out for the case that FM layer is located on top of AFM oxide layer. These oxides present unique properties, such as magnetism, superconductivity, metal-insulator transitions, electron transfer, or ferroelectricity.[5–12] In these systems, the exchange coupling affects the physical properties of these oxides and in turn some parameters of the AFM oxides can be used to manipulate the variation of EB. A representative example is the strain controls the exchange bias in FM/AFM (ferroelectric) system.[10–12] It has been reported that the EB can be influenced by many factors in FM/AFM system, such as the FM magnetization $M_{FM}$, the thickness of FM layer $t_{FM}$ or AFM layer $t_{AFM}$, and the anisotropy of AFM ($K_{AFM}$) or FM ($K_{FM}$).[2–4,13,14] The exchange bias field ($H_E$) is inversely proportional to the $M_{FM}$ and $t_{FM}$, and the $K_{AFM}$ is reported to affect the critical thickness of the AFM layer.[2,3,15,16] Furthermore, the AFM or FM domain formation is also claimed to play a dominant role in EB. Mauri et al.[17] and Malozemoff[18] predicted that $H_E \propto (K_{AFM}A_{AFM})^{1/2}/M_{FM}t_{FM}$ when the domain wall formed in AFM layer; moreover, Ball et al.[19,20] found that the domain wall might occur in the FM layer in the Fe$_3$O$_4$/NiO or Fe$_3$O$_4$/CoO systems by polarized neutron reflectometry studies, and the $H_E \propto (K_{FM}A_{FM})^{1/2}/M_{FM}t_{FM}$ depending on the domain wall formed on the FM side of the interface is also proposed,[3] where $A_{AFM}$ and $A_{FM}$ are the exchange stiffness of AFM and FM layer, respectively. Therefore, it is found that the $H_E$ can be mediated by varying $t_{FM}$, $M_{FM}$, or $K_{FM}$ for different FM materials.[2–4] To search a FM material that exhibits a big change in $K_{FM}$ leading to a large variation of the $H_E$ within the narrow temperature range will be a very interesting issue.

As one of the oldest known oxide materials, magnetite (Fe$_3$O$_4$) keeps on attracting extensive attention in fundamental science as well as for possible applications in spintronics[21–25] due to its rather unique and interesting set of electrical and magnetic properties,[26–29] and the first-order metal-insulator transition known as the Verwey transition around 124 K.[30] At the Verwey transition temperature ($T_V$), the Fe$_3$O$_4$ undergoes a structural transition from spinel cubic to monoclinic structure with a sharp change in electrical and magnetic properties.[26,27] This transition provides an external tuning capacity of the properties by varying the temperature. Therefore, it can be expected that the rapid change in $K_{FM}$ in the vicinity of the $T_V$ will result in the obvious variation of $H_E$. The exchange bias effect with Fe$_3$O$_4$ has been reported by many groups, such as the Fe$_3$O$_4$/CoO bilayers,[31–33] yet the Verwey transition is not mentioned in their systems.[31–41] Venta et al.[8] reported the effect of the Verwey transition on the exchange bias in Ni(Ni$_{80}$Fe$_{20}$)/V$_2$O$_3$ system, the Fe$_3$O$_4$ is considered to be formed with the reaction between Ni$_{80}$Fe$_{20}$ and V$_2$O$_3$, the Fe$_3$O$_4$ film is extremely thin, whereas the Verwey transition is found to disappear for very thin film (<5 nm).[42–46] Their discussion about the exchange bias affected by the Verwey transition, to our knowledge, is still debated. Therefore, it is very necessary to study the effect of the Verwey transition on the exchange bias in FM/AFM system with thicker Fe$_3$O$_4$ layer. It has been reported that the Verwey transition is greatly influenced by the thickness of Fe$_3$O$_4$ thin films.[42–46] With decreasing the thickness, the $T_V$ and the transition become lower and broader, respectively, or even disappears for very thin film (such as 5 nm in Refs. 45 and 46 or thinner than 30 nm in Refs. 42–44), thus the change in $K_{FM}$ around $T_V$ becomes very small for the thin Fe$_3$O$_4$ film. On the other hand, the $H_E$ is inversely proportional to the FM thickness $t_{FM}$.[2,3] Considering these two aspects, in order to observe a clear variation of the $H_E$ in the vicinity of $T_V$, we chose the thicknesses of Fe$_3$O$_4$ and CoO as 40 nm and 5 nm, respectively, in our FM/AFM bilayer system.

a)E-mail: xhliu@alum.imr.ac.cn

In this work, we report the exchange bias effect tuned by the Verwey transition of $Fe_3O_4$ in CoO (5 nm)/$Fe_3O_4$ (40 nm)/MgO (001) bilayer. A sharp increase in $K_{FM}$ leads to a rapid enhancement of the exchange bias field in the vicinity of $T_V$. Furthermore, compared with the $Fe_3O_4$ (40 nm)/MgO (001) thin film, the coercivity of the CoO (5 nm)/$Fe_3O_4$ (40 nm)/MgO (001) is greatly enhanced for all the temperature range, which would be induced from the strong coupling between Co spins and Fe spins across the interface, the partial interface spins of CoO rotating with magnetic field during the hysteresis loop measurement.

## II. EXPERIMENTAL METHODS

The CoO (5 nm)/$Fe_3O_4$ (40 nm) and CoO (5 nm)/$Fe_3O_4$ (20 nm) bilayers and the 40 nm-thick $Fe_3O_4$ thin film were grown on MgO (001) by using molecular beam epitaxy (MBE) in an ultrahigh vacuum system with a background pressure of $1 \times 10^{-10}$ mbar range. The 40 nm (or 20 nm) $Fe_3O_4$ was grown on clean MgO (001) using an iron flux of 1 Å per minute in an oxygen background pressure of $1 \times 10^{-6}$ mbar with a substrate temperature of 250 °C,[45] and then, 5 nm CoO was *in situ* grown on $Fe_3O_4$ (40 nm)/MgO (001) using a cobalt flux of 1 Å per minute in an oxygen background pressure of $3 \times 10^{-7}$ mbar with a substrate temperature of 260 °C. The growth temperature or the growth oxygen pressure in our work is different from that reported by Lind *et al.*[47] and Wolf *et al.*[48] To determine the structural quality and the chemical states, the films were analyzed *in situ* by using reflection high-energy electron diffraction (RHEED), low-energy electron diffraction (LEED), and X-ray photoemission spectroscopy (XPS). High-resolution X-ray diffraction (HR-XRD) was employed for further *ex situ* investigation of the structural quality and the microstructure of the films. The transport and magnetic properties of the thin films were measured with a standard four probe technique using physical property measurement system (PPMS) and superconducting quantum interference device (SQUID), respectively.

## III. RESULTS AND DISCUSSION

Figure 1 shows RHEED and LEED electron diffraction patterns of the clean substrate MgO (a) and (d), the 40 nm-thick $Fe_3O_4$ film on MgO (001) (b) and (e), and the CoO (5 nm)/$Fe_3O_4$ (40 nm)/MgO (001) bilayer (c) and (f). Sharp RHEED streaks and the high contrast and sharp LEED spots [Figs. 1(b) and 1(e)] indicate a flat and well ordered (001) crystalline surface structure of the 40 nm $Fe_3O_4$ film grown on MgO (001). The presence of the $(\sqrt{2} \times \sqrt{2})R45°$ surface reconstruction patterns both in the RHEED and in the LEED images provides another indication for the high structural quality of the $Fe_3O_4$ film.[45] The lattice parameter of $Fe_3O_4$ (8.397 Å) is nearly twice as that of CoO (4.267 Å),[49] with growing CoO (5 nm) on $Fe_3O_4$ (40 nm)/MgO (001), the spinel and reconstruction streaks of RHEED in Fig. 1(b) and corresponding spots of LEED in Fig. 1(e) disappear. The sharp RHEED and LEED patterns of CoO (5 nm)/$Fe_3O_4$ (40 nm)/MgO (001) in Figs. 1(c) and 1(f) become similar to that of the MgO (001) [see Figs. 1(a) and 1(d)]. To further check the chemical states of the CoO thin film, the CoO

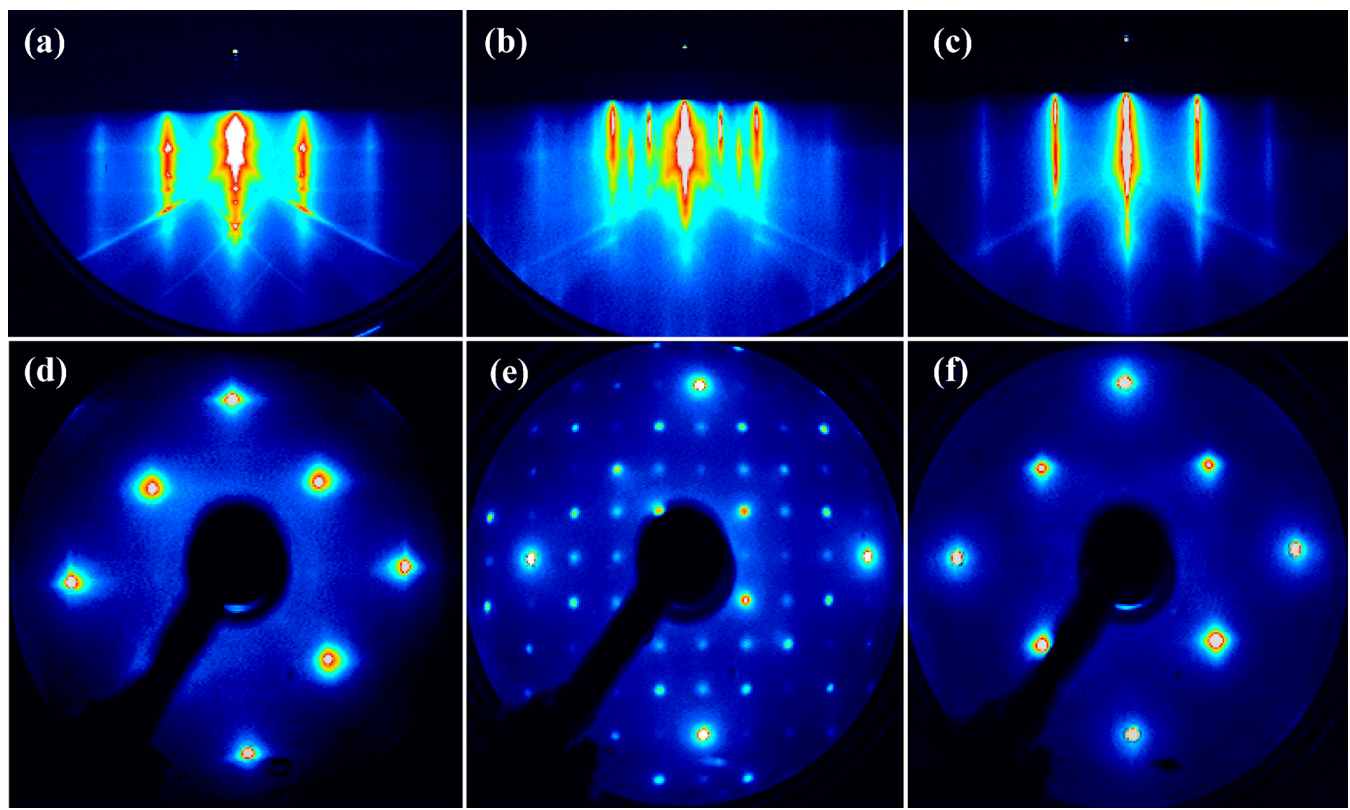

FIG. 1. RHEED and LEED electron diffraction patterns of the following: clean substrate MgO (001) (a) and (d); 40 nm-thick $Fe_3O_4$ grown on MgO (001) (b) and (e); 5 nm-thick CoO grown on $Fe_3O_4$ (40 nm)/MgO (001) (c) and (f).

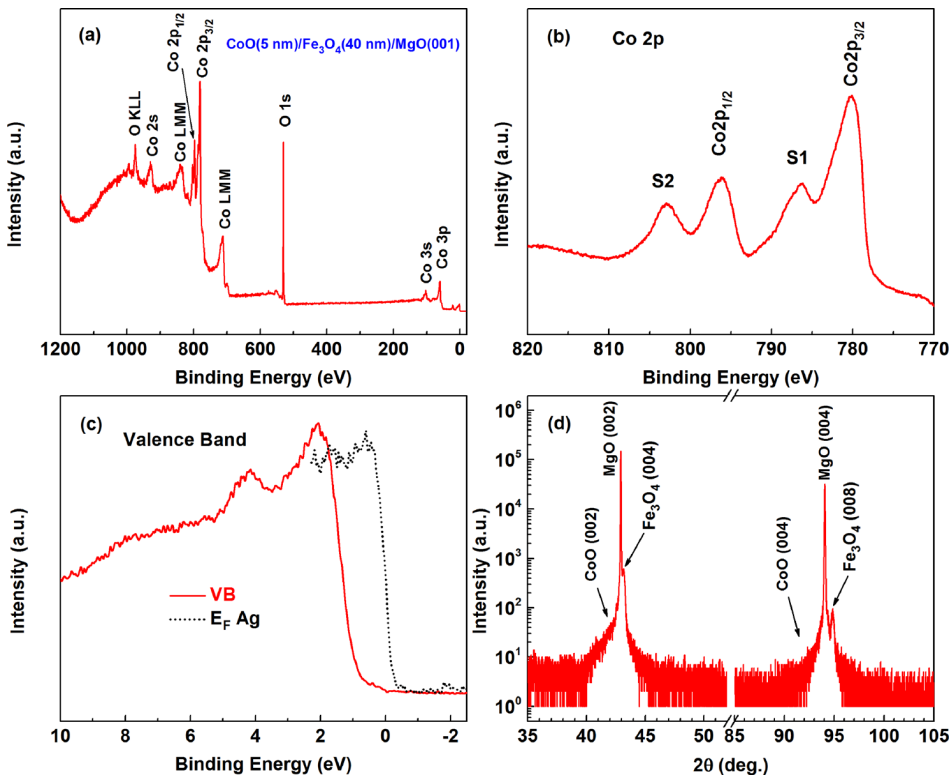

FIG. 2. XPS spectra of the following: wide scan with binding energy from 1200 to −18 eV (a), Co 2p core-level (b) and valence band (c), and long range θ–2θ X-ray diffraction pattern (d) of CoO (5 nm)/Fe$_3$O$_4$ (40 nm)/MgO (001).

(5 nm)/Fe$_3$O$_4$ (40 nm)/MgO (001) bilayer was analyzed in situ by XPS shown in Figs. 2(a)–2(c). The wide scan with binding energy from 1200 eV to −18 eV shows a clear and typical XPS pattern of CoO.[50] The Co 2p core-level XPS spectrum shown in Fig. 2(b) represents the typical characteristic for Co$^{2+}$ with a clear satellite feature at 786.3 eV and 802.9 eV, marked as S1 and S2, respectively,[50,51] and the valence band presents insulating behavior for the CoO [Fig. 2(c)]. Furthermore, the long range θ–2θ high-resolution X-ray diffraction showing only single phase of CoO (5 nm)/Fe$_3$O$_4$ (40 nm) in Fig. 2(d) also demonstrates the high quality of the epitaxial thin film.

The resistivity ($\rho$) as a function of temperature ($T$) for the Fe$_3$O$_4$ (40 nm)/MgO (001) thin film in Fig. 3(a) displays a sharp Verwey transition with a clear hysteresis [The CoO (5 nm)/Fe$_3$O$_4$ (40 nm)/MgO (001) bilayer is insulating, and we could not measure the electrical properties.] Moreover, the zero-field and field cooling (ZFC-FC) magnetization ($M$) dependent on temperature for CoO (5 nm)/Fe$_3$O$_4$ (40 nm)/MgO (001) bilayer in Fig. 3(b) shows a rapid change in magnetization at the Verwey transition, similar to that reported in the previous work.[52,53] The Verwey transition temperatures $T_{V-}$ and $T_{V+}$ are defined as the temperature of the maximum slope of log [$\rho$ ($T$)] or $M$ ($T$) curve for the cooling down and warming up temperature branches, respectively [see the insets of Figs. 3(a) and 3(b)]. The values of 114.5 K and 119.5 K for $T_{V-}$ and $T_{V+}$ [the inset of Fig. 3(a)] and very sharp transition indicate the quite high quality of our thin film, as compared to the previous work.[42–45,54–56] Furthermore, it is found that the $T_{V-}$ and $T_{V+}$ in Fig. 3(a) are consistent with those in Fig. 3(b), indicating no influence of CoO on the chemical composition of the Fe$_3$O$_4$ layer.

Magnetic hysteresis loops [$M$ ($H$)] at different temperatures of CoO (5 nm)/Fe$_3$O$_4$ (40 nm)/MgO (001) are shown in Fig. 4(a), in which each $M$ ($H$) curve was measured in the applied field of 50 kOe after field cooling ($H_{CF}$) at 10 kOe ($H$ along the [100] direction) from 300 K (higher than the Neel temperature of CoO).[2,3] Obvious exchange bias is noticed at low temperatures from 50 to 119 K, whereas it becomes very small when $T \geq 120$ K ($\sim T_{V+}$). In this system, the $H_E$ and $H_C$ are defined as $H_E = (H_L - H_R)/2$ and $H_C = (H_L + H_R)/2$, respectively; here, $H_L$ and $H_R$ are the points where the hysteresis loop intersects the field axis. Figure 4(b) shows the values of the $H_E$ calculated from Fig. 4(a) as a function of temperature for CoO (5 nm)/Fe$_3$O$_4$ (40 nm)/MgO (001) bilayer. Clearly, the $H_E$ decreases with a rising temperature and exhibits a sharp drop at $T_{V+}$. The $H_E$ is about 68 Oe at 200 K and nearly disappears at 250 K, implying the blocking temperature of CoO is about 250 K.[31–33] Similarly, the $H_E$ (T) curve for $H_{CF} = 2$ kOe also exhibits a rapid jump at $T_{V+}$ but with slightly smaller $H_E$ values due to non-fully oriented interface magnetic moments at this cooling field.[2,3] However, this sharp change in $H_E$ at $T_V$ is not observed for the CoO (5 nm)/Fe$_3$O$_4$ (20 nm)/MgO (001) bilayer [see the inset (left) of Fig. 4(b)] because of the broadened Verwey transition for the thinner Fe$_3$O$_4$ film, similar to that reported in the previous work.[31–33] For Fe$_3$O$_4$, the low-temperature monoclinic magnetocrystalline anisotropy constants are considerably greater (about 10 times) than those of the high-temperature cubic structure,[57–59] and we thus observe a great enhancement of $H_C$ at $T < T_V$, see Fig. 4(c).

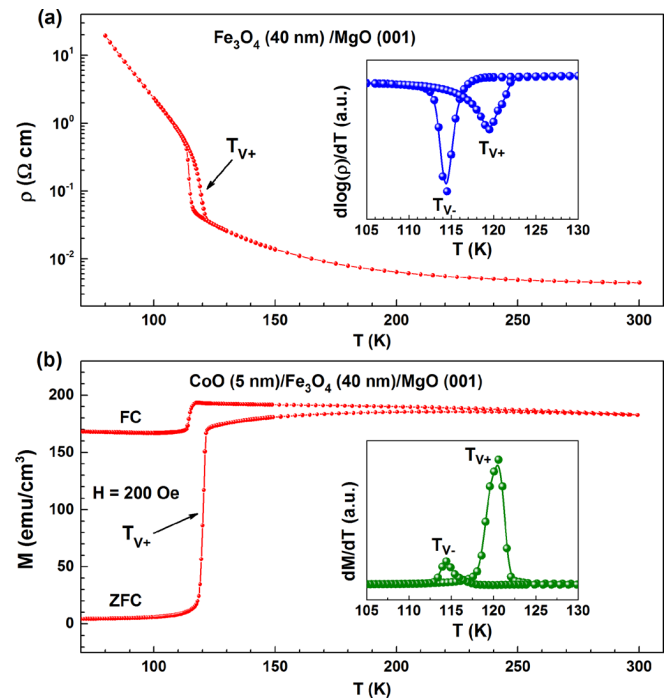

FIG. 3. Resistivity as a function of temperature for the 40 nm-thick Fe$_3$O$_4$ film grown on MgO (001) (a) and ZFC-FC temperature-dependent magnetization curve of CoO (5 nm)/Fe$_3$O$_4$ (40 nm)/MgO (001) sample in applied field of 200 Oe (b). Inset: temperature dependence of d(log$\rho$)/dT (a) and dM/dT (b) around the Verwey transition, $T_{V-}$ and $T_{V+}$ correspond to the two peaks.

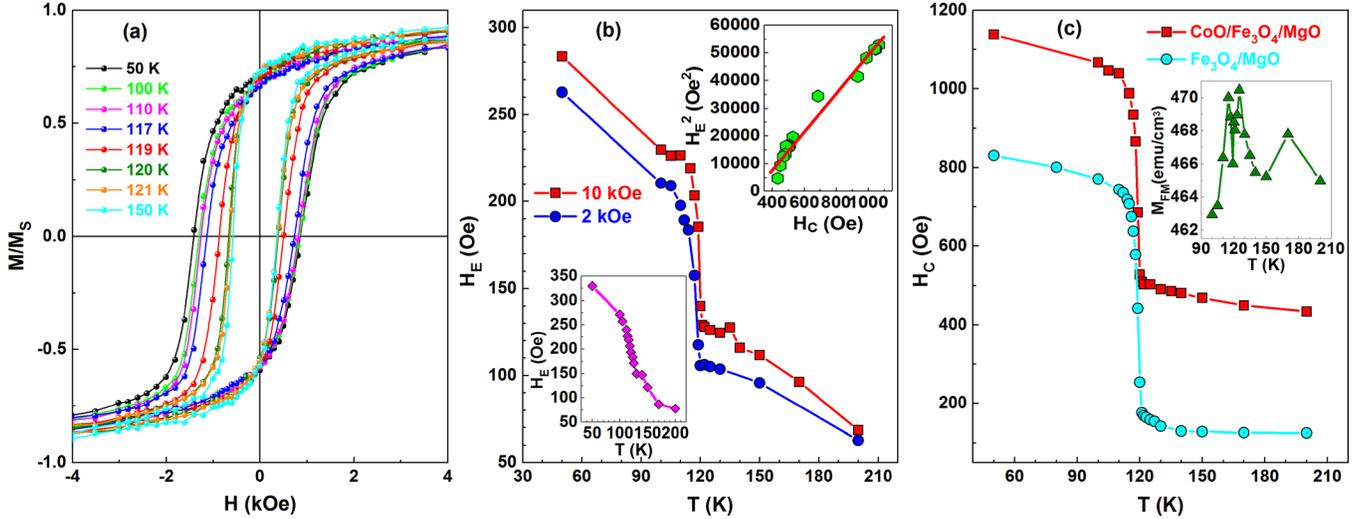

FIG. 4. Magnetic hysteresis loops at different temperatures in applied field of 50 kOe after field cooling at 10 kOe from 300 K (a) and the exchange bias field ($H_E$) as a function of temperature from 50 to 200 K at cooling field of 10 kOe and 2 kOe (b), of CoO (5 nm)/Fe$_3$O$_4$ (40 nm)/MgO (001). Inset of (b): $H_C$ vs $H_E^2$ from 100 to 200 K (right) and temperature dependence of the $H_E$ at cooling field of 10 kOe of CoO (5 nm)/Fe$_3$O$_4$ (20 nm)/MgO (001) (left); temperature dependence of coercivity field ($H_C$) for CoO (5 nm)/Fe$_3$O$_4$ (40 nm)/MgO (001) and Fe$_3$O$_4$ (40 nm)/MgO (001) from 50 to 200 K (c). Inset of (c): saturation magnetization ($M_{FM}$) as a function of temperature of CoO (5 nm)/Fe$_3$O$_4$ (40 nm)/MgO (001).

The coercivity is related to the anisotropy constants and the saturation magnetization of the FM materials and can be roughly expressed as $H_C \propto K_{FM}/M_{FM}$,[60] at the same time, the $H_E \propto (K_{FM}A_{FM})^{1/2}/M_{FM}t_{FM}$ when the domain wall formed on the FM side of the interface;[3] thus, the relationship $H_C \propto H_E^2$ can be obtained, which is well expressed as an inset (right) of Fig. 4(b), indicating the domain wall formed in the FM layer proposed by Ball *et al.*[19,20] and the tunability of the exchange bias by the Verwey transition; moreover, Ijiri *et al.*[61] also found that the CoO AFM ordering is long-range and propagates coherently through the intervening Fe$_3$O$_4$ layer in Fe$_3$O$_4$/CoO superlattices. van der Zaag *et al.*[32] calculated the low-temperature unidirectional anisotropy constant $K_E^{(0)} = \mu_0 M_{FM} H_E^{(0)} t_{FM}$, where $K_E^{(0)}$ and $H_E^{(0)}$ are the values at 0 K. Similarly, extrapolated our results to 0 K we estimated the $K_E^{(0)}$ of about 0.68 mJ/m$^2$, which is smaller than that reported by van der Zaag *et al.*[32] Furthermore, compared with the pure Fe$_3$O$_4$ (40 nm)/MgO (001) thin film, the $H_C$ of CoO (5 nm)/Fe$_3$O$_4$ (40 nm)/MgO (001) bilayer is much larger for all the temperature range [see Fig. 4(c)], meaning the strong coupling between Co spins and the Fe spins across the interface of CoO/Fe$_3$O$_4$ bilayer, that partial interface spins of CoO rotate with magnetic field during the hysteresis loop measurement.[2,3] Therefore, both the obvious exchange bias and the enhancement of coercivity are observed in our CoO (5 nm)/Fe$_3$O$_4$ (40 nm)/MgO (001) bilayer, see Figs. 4(b) and 4(c). Moreover, the $H_C$ (T) curves in Fig. 4(c) show the same Verwey transition temperature for Fe$_3$O$_4$ (40 nm)/MgO (001) and CoO (5 nm)/Fe$_3$O$_4$ (40 nm)/MgO (001), which is in agreement with that in Fig. 3. Finally, we have to point out that the saturation magnetization $M_{FM}$ also changes at $T_V$, but this $\Delta M_{FM}$ is very small,[62] only around 1% [see the inset of Fig. 4(c)]; we thus can omit the effect of $M_{FM}$ on the exchange bias in our system.

## IV. SUMMARY

In conclusion, we have investigated the exchange bias tuned by the Verwey transition of Fe$_3$O$_4$ in the CoO (5 nm)/Fe$_3$O$_4$ (40 nm)/MgO (001) bilayer. The $H_E$ is significantly enhanced because of a sharp increase in $K_{FM}$ from high temperature cubic to low temperature monoclinic structure at $T_V$. Moreover, the coercivity of the bilayer is greatly increased for all the temperature range as compared to the pure Fe$_3$O$_4$ (40 nm)/MgO (001) thin film due to the partial interface spins of CoO rotating with magnetic field during the hysteresis loop measurement.

## ACKNOWLEDGMENTS

We thank L. H. Tjeng from the MPI CPfS for stimulating discussions. This work has been supported by the Max Planck-POSTECH Center for Complex Phase Materials, the National Basic Research Program (Grant No. 2017YFA0206302) of China, and the National Natural Science Foundation of China under Project Nos. 51590883 and 51331006, and as a project of the Chinese Academy of Sciences with Grant No. KJZD-EW-M05-3.